\documentclass[conference]{IEEEtran}
\IEEEoverridecommandlockouts

\usepackage{cite,url}
\usepackage{amsmath,amssymb,amsfonts}
\usepackage{algorithmic}
\usepackage{graphicx,booktabs}
\usepackage{textcomp}
\usepackage{xcolor}

\usepackage{verbatim}
\usepackage[]{algorithm2e}
\usepackage{ifthen}
\usepackage[normalem]{ulem}
\useunder{\uline}{\ul}{}
\usepackage{pifont}
\usepackage{amsmath}
\usepackage{csquotes}
\usepackage{enumitem} 
\usepackage{listings} 

\usepackage[colorinlistoftodos,textwidth=1.22cm]{todonotes}

\def\CARONTE{{\texttt {CARONTE}}}

\def\grabsite{{\texttt{grab-site}}}

\def\BibTeX{{\rm B\kern-.05em{\sc i\kern-.025em b}\kern-.08em
    T\kern-.1667em\lower.7ex\hbox{E}\kern-.125emX}}
\begin{document}

\title{\CARONTE: Crawling Adversarial Resources Over Non-Trusted, High-Profile Environments}

\author{\IEEEauthorblockN{Michele Campobasso}
\IEEEauthorblockA{
\textit{University of Bologna, IT}\\
michele.campobasso2@studio.unibo.it}
\and
\IEEEauthorblockN{Pavlo Burda}
\IEEEauthorblockA{
\textit{Eindhoven University of Technology, NL}\\
p.burda@tue.nl}
\and
\IEEEauthorblockN{Luca Allodi}
\IEEEauthorblockA{
\textit{Eindhoven University of Technology, NL}\\
l.allodi@tue.nl}
}

\maketitle

\begin{abstract}
The monitoring of underground criminal activities is often automated to maximize the data collection and to train ML models to automatically adapt data collection tools to different communities. On the other hand, sophisticated adversaries may adopt crawling-detection capabilities that may significantly jeopardize researchers' opportunities to perform the data collection, for example by putting their accounts under the spotlight and being expelled from the community. This is particularly undesirable in prominent and high-profile criminal communities where entry costs are significant (either monetarily or for example for background checking or other trust-building mechanisms). This paper presents \CARONTE, a tool to semi-automatically learn virtually any forum structure for parsing and data-extraction, while maintaining a low profile for the data collection and avoiding the requirement of collecting massive datasets to maintain tool scalability. We showcase the tool against four underground forums, and compare the network traffic it generates (as seen from the adversary's position, i.e. the underground community’s server) against state-of-the-art tools for web-crawling as well as human users. 
\end{abstract}

\begin{IEEEkeywords}
underground, stealth monitoring, data collection, high-profile communities
\end{IEEEkeywords}

\section{Introduction}

The monitoring of underground activities is a core capability enabling law enforcement actions, (academic) research, malware and criminal profiling, among other activities. Currently, monitoring activities focus on the rapid collection of massive amounts of data~\cite{soska2015measuring}, that can then be used to train machine learning (ML) models to, for example, extend available parsing capabilities to different forums or underground communities. Indeed, the proliferation of underground criminal communities makes the scalability of monitoring capabilities an essential aspect of an effective, and extensive, data collection, and ML has been the clear go-to solution to enable this. However, this comes at the high price of having to collect large volumes of data for training, raising the visibility of the researcher's activity and interest in the criminal community.

Indeed, the scientific literature showed that not all communities are born the same~\cite{Herley-2010-EISP}; on the contrary, the majority of underground communities appear largely uninteresting (even when generating massive amounts of data about alleged artifacts~\cite{van2018plug}), both in terms of economics and social aspects~\cite{Allodi-TETCS-15,whyforums}, as well as in terms of (negative) externalities for society at large~\cite{soska2015measuring,Nayak-2014-RAID}. Whereas there are only a limited number of `interesting' communities to monitor, gaining access to these may be less than trivial in many cases, particularly for forum-based communities and markets~\cite{whyforums,allodi2017economic}: high entry costs in terms of entry fees, background checks, interviews, or pull-in mechanisms are becoming more and more adopted in the underground as a means to control or limit the influence of `untrusted' players in the community~\cite{whyforums,allodi2017economic}.
Under these circumstances, researchers and LE infiltrating underground communities may face significant opportunity costs whereby increasing monitoring activity may also jeopardize their ability to monitor the very community(-ies) in which they wish to remain undercover: network logs and navigation patterns of crawling tools (authenticated in the communities using the researcher's credentials) can put the real nature of that user's visits under the spotlight, and lead to blacklisting or banning from the community. This is particularly undesirable in high-profile communities where the cost of re-entry can be high.

\begin{figure}[t]
\centering
\includegraphics[width=0.7\columnwidth]{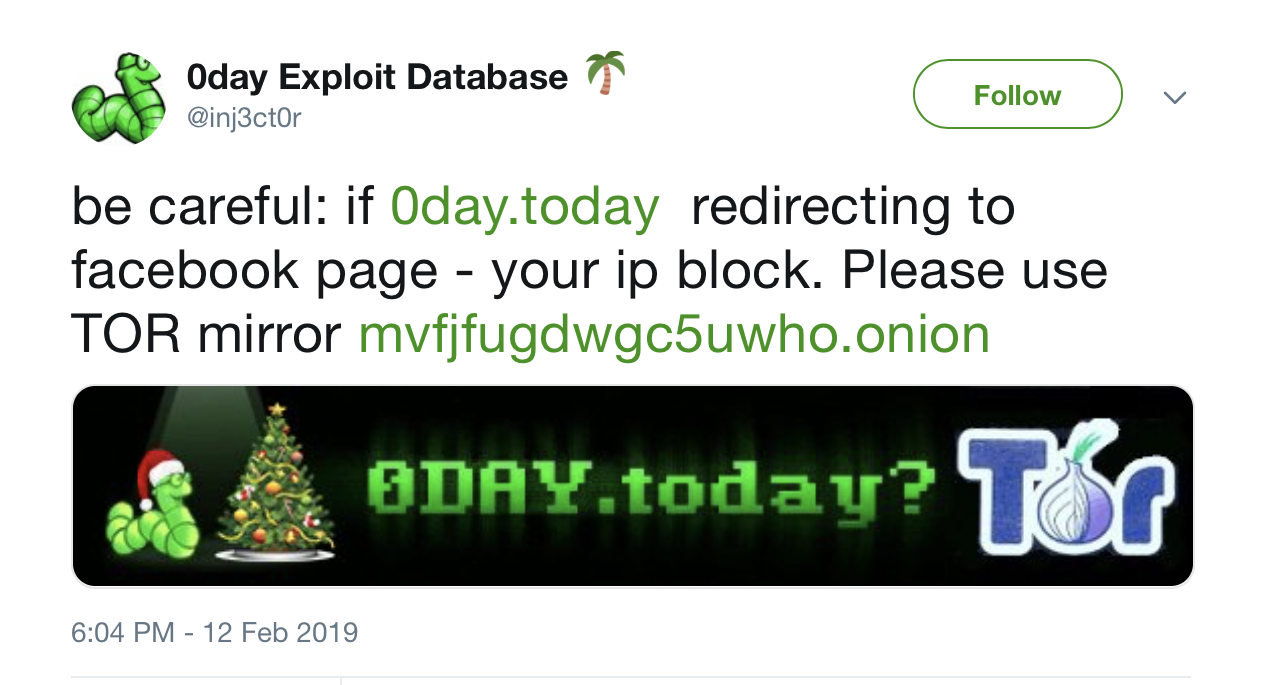}
\label{fig:0daytoday}
\caption{Example of inbound traffic monitoring from criminal communities}
\end{figure}

Anecdotal evidence shows that monitoring incoming traffic, for example for robot detection or source-IP checking, is a countermeasure that underground communities may employ to limit undesired behaviour. Some communities explicitly acknowledge the adopted countermeasures (see for example Figure~\ref{fig:0daytoday}), 
others explicitly state that they are aware of the monitoring operations of LE and other `undesirable' users; for example, the administrator of one prominent underground forum for malware and cyber-attacks that the authors are monitoring, states explicitly: \textit{``Forums like this are being parsed by special services and automatically transfer requests to social network accounts and e-mails.''} This significantly inhibits researchers' ability to build scalable, reusable parsing modules, as the collection of large amounts of data to train the associated ML algorithms may be slow or carry significant risks of exclusions from the monitored communities. Pastrana et al. \cite{pastrana2018crimebb} lead the way in identifying \emph{stealthiness} as a requirement for systematic underground resource crawlers, with many recent works not explicitly mentioning these aspects~\cite{portnoff2017tools,lai2011automatic}.

In this work we present \CARONTE, a tool to monitor underground forums that: (1) can be configured to semi-automatically learn virtually any forum structure, without the need of writing ad-hoc parsers or collecting and manually classify large volumes of data; (2) implements a simple user model to mimic human behaviour on a webpage, to maintain a low profile while performing the data-collection. We showcase the tool against 4 underground forums, and compare the network traffic it generates (as seen from the adversary's position, i.e. the underground community's server) against state-of-the-art tools for web-crawling. Our results clearly show that both \CARONTE's request patterns as well as the completeness of the downloaded resources page are significantly closer to humans when compared to other SoA crawling tools.

This paper proceeds as follows. In Section~\ref{sec:background} we discuss relevant background and related work; Section~\ref{sec:caronte} presents the tool, whereas Section~\ref{sec:expvalidation} presents the experimental validation and results. ~\ref{sec:discussion} deepens the impact and limitations of \CARONTE{}, and Section~\ref{sec:conclusions} concludes the paper.

\section{Background}
\label{sec:background}

Cybercrime monitoring has mainly been implemented through \textit{ad-hoc} tools to scrape adversarial platforms that don't scale up with the number of sources and the variety of the content; nonetheless, most of them were more concerned on developing techniques that enable underground economy discovery, key hacker identification \cite{abbasi2014descriptive} and threat detection \cite{benjamin2015exploring}, disregard stealthiness in favour of parsing volumes~\cite{lai2011automatic,portnoff2017tools}, with few notable exceptions~\cite{pastrana2018crimebb}. In this section we discuss the changing threat model against which these solutions are used, and the technical means by which their employment can be detected by adversaries.

\subsection{A changing threat model for cybercrime monitoring}

Cybercriminals have demonstrated to be increasingly aware of the mounting interest from scientific and nation state sponsored investigations, pushing them to start developing techniques to avoid unwanted actors and data gathering in their communities \cite{oerlemans2017investigating}. Retaliation activities on the side of the cybercrooks have made the news in the past~\cite{whyforums,allodi2017economic,Allodi-TETCS-15}, including threatening journalists~\footnote{A man accused of trying to frame a blogger with heroin is in big trouble, Business Insider. Visited April 2019. \url{https://www.businessinsider.com/brian-krebs-heroin-threat-hacker-extradited-2015-10/?international=true&r=US&IR=T}} and academic researchers~\cite{hine2017kek}. The attention posed by cybercriminals to the public sphere is also reflected on the technology and administrative procedure they employ to detect or stop possible `intrusions' in their communities. Part of these techniques are centered on the evaluation of a perspective member at the act of registration on these platforms~\cite{whyforums,Allodi-TETCS-15}, or on the continuous monitoring of community participation by each member~\cite{allodi2017economic}. Similarly, recent evidence suggests cybercriminals may be monitoring and auditing the traffic on the web servers, both to prevent access from undesired IP ranges (e.g.~see Figure~\ref{fig:0daytoday}), and to mitigate external threats such as denial of service attacks. These technologies identify patterns and anomalies in network traffic to detect undesired activities, requests generated by robots and crawlers and, if necessary, take action to limit those~\cite{qassrawi2010client,fallmann2010covertly}. 
Whereas researchers can build profiles to go undercover in certain communities, thereby passing the first access filter enforced by cybercrime community entry regulations~\cite{allodi2017economic}, network monitoring operations remain an unmitigated threat to automation of data collection once inside the forum, particularly at the face of an evolving adversary (i.e.~ the cybercrooks behind the platform).

\subsection{Robot and crawler detection}

With the birth of the big data society, crawling has become a conspicuous portion of the Internet traffic \cite{bai2014analysis} and an unwanted practice from website owners, due both to network resource consumption and to the lack of an explicit permission to a third-party to massively download all the website content for unknown goals, often resulting in a privacy violation \cite{zhang2013novel}\cite{ford2004googling}. 

Several anti-crawling techniques have been developed; a great number of these are based on minimal traffic patterns analysis, often focused on monitoring characteristics of HTTP traffic observable from logs. The monitored characteristics include the rate of requests, the length of browsing sessions, lack of cookie management, presence of bogus user agent, JavaScript execution, access to \textit{robots.txt} file, usage of HEAD HTTP requests, cherry-picking of requested resources and the lack of a referrer in HTTP requests  \cite{zhang2013novel}\cite{sardar2014detection}\cite{doran2013comparison}. 
These strategies are quite simplistic and don't provide consistent and reliable crawler detection, ignoring the chance that a focused and stealthier crawler can act in disguise, tampering with information in the requests. Since our adversary could legitimately have advanced skills in computer matters, these requisites remain relevant, but aren't sufficient to keep a robot undercover. Additional efforts have been made for creating more reliable methods; the state-of-art techniques for detecting automated activity on a website include pattern recognition, like loopholes detection and breadth first or depth first strategies, JavaScript fingerprinting and tracking, and Turing tests, on top of other strategies \cite{kwon2012web2}\cite{doran2011web}\cite{von2003captcha}. Turing tests as CAPTCHAs can be outsourced at extremely low prices \cite{prowebscraper_2018} or solved via OCR \cite{korakakis2014automated} and the production of not suspicious traffic can be obtained with a focused crawler that acts with some precautions. Moreover, the context of our studies brings us to platforms in the dark-web where Turing tests that require JavaScript enabled are almost non existent, due to linked risks (e.g. allowing in the past to bypass completely the anonymization of Tor \cite{torjs_thehackernews_2013}\cite{tornoscript_securityweek_2018}).

Zhang et al. \cite{zhang2013novel} propose a dynamic blocker for crawlers analyzing some traffic patterns, such as the complete exploration of the resources linked to a page (attachments, links, ...), the nonacceptance of cookies, bogus user agents in HTTP requests and high fetch rates. Stevanovic et al. introduces some additional checks compared to the previous analyzed works, such as the \textit{HTML/image ratio}\footnote{HTML/image ratio is a numerical attribute calculated as the
number of HTML page requests over the number of image file
(JPEG and PNG) requests sent in a single session.}, which tends to be very high for crawlers, the \textit{number of PDF/PS file requests}, the percentage on total requests of answers with 4xx error codes and unassigned referrers, which show high scores for robots \cite{stevanovic2012feature}. 
Doran et al. propose to recognize crawlers in real-time \cite{doran2011web}. In particular, their work they provide a 4-tier analysis based on \textit{Syntactical log analysis}, \textit{Traffic pattern analysis}, \textit{Analytical learning techniques}, and \textit{Turing test systems}. For what concerns more behavioral patterns, Kwon et al. have studied how crawlers generally have a \textit{monotonous behavior} in the type of requested resources. Their attempt therefore is to classify crawlers based on the \textit{"switching factor"} between text and multimedia contents \cite{kwon2012web2}. Other studies regarding behavioral patterns are by Balla et al., who analyze the time between one request and another and when these are issued (like during night time, making it more suspicious) \cite{balla2011real}.

Crawler detection patterns aside, data gathering, ready to use and well structured, is not an easy task to accomplish. Forum structures may vary a lot, depending on the forum platform adopted and their configuration. In particular, the goal is not to scrape the entire pages to dump them on disk, but to extract and structure data for further analysis; for this reason, the crawler need to be instructed on what resources are required to be collected, how to reach them and what do they mean. Therefore, a knowledge base should be created for the crawler in a reliable way, enabling the forum traversal in the required areas through the identification of the existing resources of each page of interest.

\subsection{Modeling `regular' user behaviour}

Studies on user browsing behaviour broadly distinguish between \emph{click patterns} and \emph{time patterns}. 

\emph{Click patterns}. Click models are used to evaluate user decisions in considering a topic or hyperlink relevant to the specific purpose of their navigation or query \cite{dupret2008user}. Derived approaches consider \emph{single-browsing} and \emph{multi-browsing} models to infer user behaviour as a function of the purpose of the navigation, in particular distinguishing between \emph{navigational} and \emph{informational} queries, whereby the user wants to reach a specific resource (likely producing one click at a time), or is interested in exploring new information (likely producing multiple clicks at a time)~\cite{dupret2008user,guo2009efficient}. These models show that past behaviour or user interest are useful predictors of which clicks will happen in the future~\cite{dupret2008user}. In our context, forum-browsing clearly covers both dimensions, depending on whether the user aims at retrieving specific information (e.g. updates in a thread of previous interest to the user), or to explore the content of a forum section.

\emph{Time patterns}. More broadly, these dynamics are explained in the information retrieval literature as dependent on the user's task~\cite{baeza2011modern}. The decision of a user to click on a specific resource depends on its perceived and intrinsic relevance w.r.t. the user's goal, and is bounded by how many topics need be opened to find the answer the information the user is interested in~\cite{dupret2010model}. Post-click user behaviour (i.e. what the user does one he or she reaches the clicked resource) has been shown to be directly related to the relevance of the document~\cite{guo2012beyond}. Post-click behaviour includes variables associated with mouse movements, scrolling, and eye-tracking~\cite{guo2012beyond,dupret2008user}, clearly showing that what the user does, and how much time the user spends on a webpage, varies as a function of the relevance of the webpage. Indeed, a user's \emph{inaction} on a webpage has been shown to be relevant to model the quality of dynamic systems such as recommendation systems~\cite{Zhao-inaction-2018}. Part of that behaviour can be quantified by considering how quickly users can be expected to process the relevant information~\cite{averagespeed}. Data around this subject is scarce and quite diverse; some sources refer the average reading speed to be around 200-250WPM (Words Per Minute) with a comprehension rate of 50/60\%~\cite{averagespeed}, others report that for reading some technical content with a good proficiency, the speed can be around 50-60WPM.

\section{\CARONTE}
\label{sec:caronte}

\subsection{Design}
From the literature analysis in the previous section, we derive a set of desiderata for \CARONTE. 

\subsubsection{Functional and behavioural requirements}

First and foremost, \CARONTE\ must be able to semi-automatically learn forum structures without the need for extensive pre-collected datasets on which to train automated models~\cite{portnoff2017tools}. This should be a one-time only process, employed for each new forum structure that has not already been learned.
Further, \CARONTE\ must have the ability to diverge from crawler behavior and, where possible, to mimic human behavior. In this regard, as emerged from the time patterns paragraph, keeping in mind that one significant aspect of crawlers is their greed in resources, \CARONTE\ shouldn't exhibit high fetch rates and mimic as much as possible human's time to browse and read resources, whether the content is appealing for it or not. Therefore, we model \CARONTE\ to mimic interest to a specified subset of the forum, exploring only certain sections of it, accordingly to the hypothetical goals of our modelled actor. The interested topics and sections of the forum will be taught during the learning phase. Then, \CARONTE\ will be able to receive instructions about which areas are valuable to crawl and which to skip. The forum contents will be explored both through \emph{navigational} and \emph{informational} queries; in particular, \CARONTE\ will access quickly  resources, like posts in threads already read and the resources related to path traversing that occur from the landing page to the section of interest, while it will take more time and produce less frequent clicks while staying on pages with new content from the section of interest. To improve its stealthiness on this aspect, we design a navigation schedule on a forum like an actual human being having in mind variables such as time of  day and stochastic interruptions.

\subsubsection{Technical requirements}

To avoid detection at the network level, \CARONTE\ will have to act indistinguishably to a regular browser in terms of generated traffic and differing to regular crawlers. The primary aspect is to produce not suspicious HTTP requests against the webserver; crawlers' traffic is characterized by the adoption of HEAD HTTP requests to determine whether the resource to download is of their interest or not, non-filling of referral link field in requests and by the usage of a bogus user agent \cite{sardar2014detection,doran2011web,stevanovic2012feature,jacob2012pubcrawl,balla2011real}. Further, crawlers might be interested on fetching only text content, refusing to download styles, images \cite{stevanovic2012feature} and JavaScript, (e.g. to minimize network footprint) or won't actively execute client-side code such as JavaScript, handle sessions and cache as a 'regular' web browser would do. In our study case, we assume legitimate to have JavaScript completely disabled for increasing the anonymity of our tool; this countermeasure inside the Dark Web is quite common and shouldn't raise any suspect.

With this in mind, we identify the need of a fully functional browser that by design covers all these aspects coherently with a legitimate one, but that offers the possibility to be maneuvered programmatically. Table~\ref{tab:requirements} provides an overview of the identified requirements for \CARONTE.

\begin{table*}[t]
\centering
\small
\caption{Summary of identified requirements for \CARONTE}
\label{tab:requirements}
\begin{tabular}{p{0.15\textwidth}p{0.25\textwidth}p{0.5\textwidth}}
\toprule
Requirement & Description & Implementation\\
\midrule
Learning forum structures & Understanding forum structure, how to browse it and where valuable information is & Creation of a human supervised learning module that identifies needed resources\\

Regular browser behaviour & Realistic user agents, caching behaviour, referral handling & Exploration of required sections only, throttling requests accordingly to text volume of the page, mimicking reading time. Confining crawling activity in semi-random time slots during the day and suspending it for random amounts of time during the day\\

Realistic browser configuration & Addons install and pages download feature &Install NoScript and Page Save WE, preparation of the browser to support shortcuts for downloading a page\\

Anonymity & Browsing session needs to be anonymous & TOR Browser adoption, JavaScript disabled at browser level and changing default to refuse JavaScript and active content\\

\bottomrule
\end{tabular}
\end{table*}

\subsection{Architecture and implementation}

\begin{figure}[t]
\centering
\includegraphics[scale=0.45]{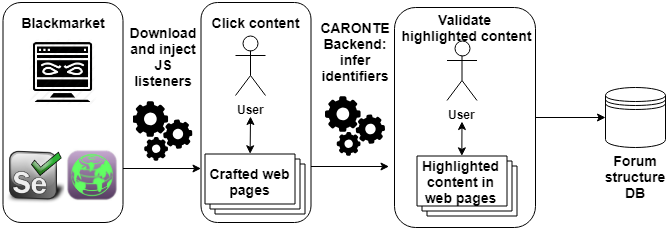}
\caption{CARONTE trainer module structure.}
\label{fig:trainerstructure}
\end{figure}

\CARONTE\ adopts a two-tier architecture, separating the \emph{training} from the \emph{crawling} operations.

\subsubsection{Trainer module}

\paragraph{Base mechanism}

\begin{figure}[t]
\centering
\includegraphics[scale=0.32]{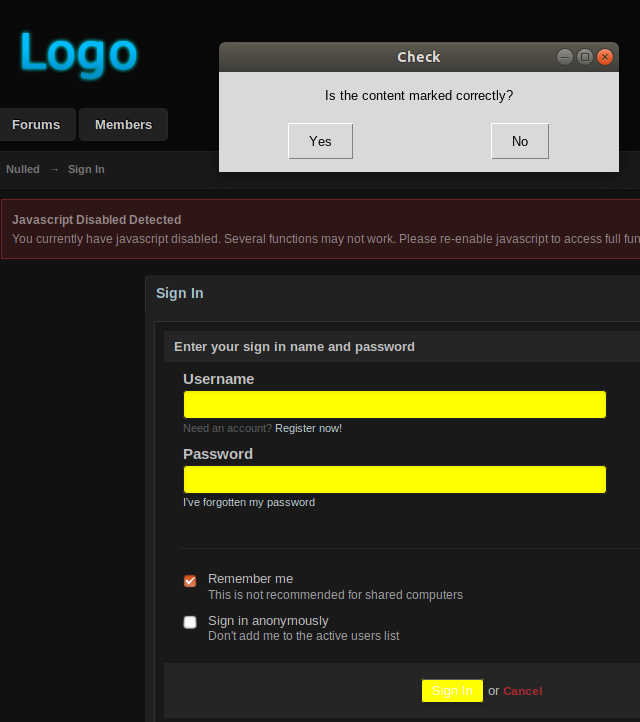}
\caption{Validation of identifiers inferred.}
\label{fig:identifiers}
\end{figure}

\begin{figure}[t]
\centering
\includegraphics[scale=0.265]{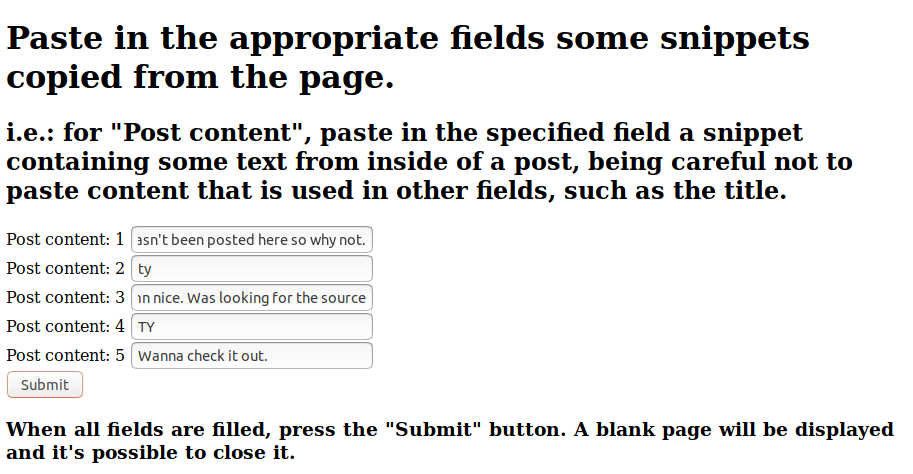}
\caption{Gathering of text snippets from the saved page (in next tab).}
\label{fig:gathering}
\end{figure}

The trainer module has the task to build a knowledge base for traversing the forum structure \ref{fig:trainerstructure}. For each page where relevant content or fields are present, the trainer will load, save and render a modified copy of it to the user. For each of them, the operator will be asked to click on the desired resources inside of the rendered page. Before being rendered, pages are preprocessed; in particular,  we inject JavaScript scripts to allow \CARONTE\ to gather the events triggered by the human operator. With different combinations of \texttt{onclick()} and \texttt{addEventListener()}, we control these interactions and generate AJAX requests against \CARONTE's backend. The payload of these requests is a resource identifier (see "Resource identifiers" paragraph in this section) that will allow the crawler module to access to the required information or interact with it, where necessary. Subsequently, it then proceeds to render again the saved page, but highlighting the previously identified content, allowing the user to confirm if the identifiers for the resources have been inferred correctly by the tool or not (Figure~\ref{fig:identifiers}).
In some cases user-generated clicks are not possible or we aim to identify a group of resources. For example, this is the case for identifying multiple posts inside of a thread; for this kind of resources, our goal is to infer a resource identifier that can operate like a regular expression, enabling the tool to resolve all the required elements on the page. 
Our strategy here is based on the collection of multiple snippets of text contained in each of these resources (Figure~\ref{fig:gathering}). For each of the received fragments, \CARONTE\ will query the JavaScript engine embedded in the browser handled by Selenium in order to resolve their identifiers and, through syntactical similarity, generate a matching one. Text content will be gathered with the help of the human operator in a special page (here referred to as \textit{content collector page}) that is presented to the user together with the original page.

\begin{figure}[t]
\centering
\includegraphics[scale=0.55]{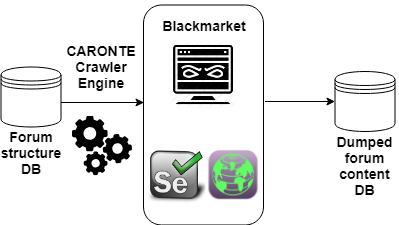}
\caption{CARONTE crawler module structure.}
\label{fig:crawlerstructure}
\end{figure}

\paragraph{Resource identifiers}
The desired resources can be identified through two different approaches: \textbf{XPath} or \textbf{HTML common classes}. XPath is a standardized query language that identifies elements inside of a XML-like document; it supports regular expressions for matching several elements. HTML classes instead are attributes assigned to nodes of an HTML file for which different styles are assigned. Even though XPath is an \textit{ad-hoc} technique for identifying elements in a HTML page, sometimes inferring HTML classes is easier than XPaths. During the training phase, when a resource is clicked, the loaded page will identify the associated identifier through a series of heuristics, and send it to the backend. If the resources are multiple, the \textit{content-collector page} will be rendered with the downloaded page and the user will fill the fields with the required data. The process to identify the most likely resource identifiers depend on the data structure and the number of classes associated with that resource. \CARONTE\ supports the following four cases: 
\begin{itemize}
    \item \textbf{Technique 1}. Extract the XPath of the exact resource. If the resources are multiple, the most frequent XPath will be the candidate;
    \item \textbf{Technique 2}. Extract XPath of the exact resource, but the last node is truncated. The XPath approach may fail due to the presence of extra HTML tags (e.g. due to text formatting), that can then be disregarded. If the resources are multiple, the most frequent XPath will be the candidate after removing the last node;
    \item \textbf{Technique 3}. The class of the exact resource. If the resources are multiple, the most frequent class will be the candidate. This approach solves the problem of calculating an XPath in a page where the content is dynamic, resulting in a non predictable XPath for a certain resource, depending on the loaded content in the page. If the resources are not assigned to a class, the element will be replaced with its parent, which will act as a wrapper;
    \item \textbf{Technique 4}. Two classes of the exact resource. If the resources are multiple, the two most frequent classes will be the candidates. This approach is adopted to handle elements in a page that exhibit the same class of the desired content, resulting in a misclassification. Therefore, this strategy allows to have a stricter condition on the searching criteria for the required resource. If the resource(s) has no class, the element will be replaced with its parent, which will act as a wrapper. 
\end{itemize}

\subsubsection{Crawler module}

Based on the structural details collected with the trainer module, the crawler module will traverse the forum to reach the required resources, explore threads and collect all the required data. The crawler will also embody the requirements of being compliant with the traffic generated from a regular browser while camouflaging its nature adopting low fetch rates for pages. How time is calculated before accessing to the next resource is deepened in the Reading Time paragraph under Implementation section.

\CARONTE\ further keeps track of updated threads and selects those opportunistically for visiting. Threads that have not been updated are not traversed a second time.

\subsection{Behavioral aspects}

\subsubsection{Legitimate browser traffic - Browser}

To implement \CARONTE's browser functionalities we adopt \textbf{Tor Browser Selenium}, or \textit{tbselenium} for short. \textit{tbselenium} accesses \textit{geckodriver}, the browser engine branded Mozilla that allows to maneuverer the browser's behavior and UI. Moreover, \textit{tbselenium} exposes an interface for customizing the environment and, finally, produces traffic identical to Tor Browser.

\subsubsection{Mimicking legitimate human traffic}

\paragraph{Work schedule}

\CARONTE\ can be configured to work within pre-defined timeslots during the week or in the weekends, late afternoons and evenings during the week and all three sessions on weekend. Between each session, a randomized time of inactivity simulates short pauses (between 5 minutes and half an hour) and longer ones at pre-defined times (e.g. 2 hours around dinner time). These can be configured. Each session has a start time and an end time; each of them can vary of up to 25\% of the total duration of the crawling session randomly. Each session has the 20\% of chance to be skipped. Nonetheless, we would avoid to have 24 hours of inactivity, so if there's no sessions scheduled in the next 24 hours, a compatible one with the default schedule will be executed. Start and end times are shifted accordingly to the timezone of the geographical location of our forum user profile.

\paragraph{Reading time}

The time spent between two requests is calculated according to two main criteria: 
\begin{itemize}
	\item If the current page doesn't show significant content to be read (e.g. pressing login button, reaching the section of interest of a forum, moving to page 2 of a forum section, ...) or the content has been already read (a thread may contain new messages, therefore old will be skipped), the time spent before going to the next page is a random number of seconds between 3 and 7. This decision is based on the fact that the information on the page is more essential and visual. This enables our fake actor to skim rapidly and choose what to read, resulting to fulfill the expectation of having a \textit{navigational queries} pattern;
	\item If the current page is the body of a thread, the tool will wait, for each unread post, an arbitrary amount of seconds calculated as the time to read the post at a speed in the range of 120-180 WPM. This behavior validates the expectation of producing \textit{informational} queries.
\end{itemize}

\paragraph{User event generation} \CARONTE{}'s modeled user goal is to reach the threads of interest and iterate them to extract their content. When starting the crawling process, \CARONTE\ loads the forum homepage, as it was typed on the address bar, then reaches the login page. Once logged in, it reaches one of the sections of interest expressed during the training and opens a thread per time (if it has been never read or has new replies). For each thread, it browses each page until the thread has been read in the whole. The click patterns generated match the purpose of our fake user, which considers relevant the content of pages with a significant quantity of text like a thread instead of a login page.

\section{Experimental validation}
\label{sec:expvalidation}

\subsection{Forum selection}

In order to proof \CARONTE's capabilities against different forums, we selected four real-world criminal forums built on top of different platforms. The candidates (Table~\ref{tab:forums}) correspond to a consistent representation of the most common forum platforms wildly adopted on the Web \cite{abbasi2014descriptive} \cite{benjamin2015exploring} \cite{portnoff2017tools} \cite{soska2015measuring} \cite{blacktds}. 

We first reproduced four live hacker forums by scraping them and hosting their content on a local server at our Institution. Before reproducing the content on our systems we inspected the source code and scanned it with \texttt{VirusTotal.com} to assure malicious links or code was not present.
Forum mirrors include multimedia content, styles and JavaScript. 
To avoid provoking misservice on the server side while scraping the forums, we avoided aggressive scraping. As our interest is to have an appropriate test-bed to evaluate \CARONTE's overall performance, the nature (or quality) of the content of the forums is irrelevant for our purposes.

\begin{table*}[t]

\centering
\small
\caption{Scraped forums for our testbed.}
\label{tab:forums}
\begin{tabular}{llll}
\toprule
Forum                         & Time span                 & Forum software          & Obtained with \\ 
\midrule
https://nulled.io                      & 14 Jan 2015 - 06 May 2016 & IP Board 3.4.4        & Online dump \\ 

http://offensivecommunity.net & Jun 2012 - 6 Feb 2019     & MyBB (unknown version) & HTTrack 3.49.2 \\ 

http://darkwebmafias.net                & Jun 2017 - 7 Feb 2019     & XenForo 1.5           & A1 Website Downloader 9 \\ 

http://garage4hackers.com        & Jul 2010 - 4 Feb 2019     & vBullettin 4.2.1      & A1 Website Downloader 9 \\ 
\bottomrule
\end{tabular}
\end{table*}

\subsection{State-of-art tools selection}
To provide a comparison of \CARONTE{}'s capabilities against other tools, we select three among the available ones:
\begin{itemize}
    \item A1 Website Download: shareware crawler specialized in downloading forum content. Through a fine-grained customization wizard, it is possible to use configuration presets that fit better the crawling process against a certain forum software, optimizing its performances;
    \item HTTrack: probably the most famous tool for downloading websites, HTTrack provides several tweaking features through regular expressions for downloading a website;
    \item \textit{grab-site}: fully open-source, grab-site is a regular crawler for downloading large portions of the web, powered by the Archive Team. 
\end{itemize}

\subsection{Training phase}

The approach adopted by \CARONTE\ to discover the structure of a forum has proven effectiveness over our tests. 
In order to get the structure of a forum, we rely on the predictability of the structure of a forum in the future in terms of XPath and HTML classes. This is true in the majority of the cases; from the literature analysis and empirical evaluations of the most common forum structures \cite{cai2008irobot} \cite{lim2013generalized} \cite{zhang2018idetector}, we found no evidence of dynamically-loaded forum structures that would alter the DOM structure at each visit or while being on a page. This seems well in line with environments like the Dark Web, where platform simplicity and functionality, as well as predictability, are desirable~\cite{whyforums}. During this phase, all the countermeasures that disable the download and execution of active content and JavaScript are in use as well.

\subsubsection{Problems and solutions}
\emph{Post details mismatch avoidance.} We've found out that seldom post details like authors and date have different structural identifiers or are displaced differently, causing the crawler module to miss them and associate a post's details to another, due to the different cardinality of the identified ones (from the crawled posts, in our database we notice that only 8 author names have been missed in 12854 posts for IP Board 3.4.4). Even though \CARONTE\ is not able to recover them, we model the post as an \textit{unique container} (which we call \textbf{post wrapper}) where details are anchored to it. By doing so, the identifiers are calculated relatively to the post and we thus avoid accidental post assignment to wrong ids. 

\begin{figure}[t]
\centering
\includegraphics[scale=0.55]{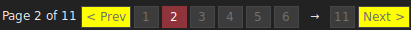}
\caption{Resource name collision.}
\label{fig:collision}
\end{figure}

\emph{Inconsistent reference to navigation button in forums.}
Depending on the forum platform and on the adopted configuration (e.g. forum skins or themes), HTML tags may have different names and usages. For example, vBullettin 4.2.1 and XenForo 1.5 adopt the same HTML tag id or class for both the forward navigation button and back inside of a thread or section of a forum. During the training stage of \CARONTE\, inferring the class of this element leads to the unwanted result of identifying both buttons with the same rule (Figure~\ref{fig:collision}). 
This would result in moving back and forth between the first and the second page. \CARONTE\ manages this issue in the training phase by loading the next page and asking the user if the highlighted part of the DOM is one element only, or more. In the second case, \CARONTE\ keeps record of the conflict and accesses the second retrieved element when this case occurs.

\subsubsection{Training evaluation}

\begin{table}[t]
\small
\centering
\caption{Treatment combination and experiments.}
\label{tab:identifiers}
\begin{tabular}{p{0.35\columnwidth}p{0.09\columnwidth}p{0.09\columnwidth}p{0.09\columnwidth}p{0.09\columnwidth}}
\toprule
Forum                         & Exact XPath & Parent XPath & Single Class & Double Class \\ \midrule
nulled.io             & 10          & 1            & 2            & 0            \\ 
offensivecommunity.net & 9           & 2            & 2            & 0            \\ 
darkwebmafias.net      & 9           & 2            & 2            & 0            \\ 
garage4hackers.com     & 8           & 1            & 3            & 0            \\ 
\bottomrule
\end{tabular}

\end{table}

Depending on the peculiarities of the forum against which \CARONTE\ has been trained, different strategies have been adopted to determine the resource identifiers ("Resource identifiers" paragraph). A summary about the identification strategies used per forum is available in Table~\ref{tab:identifiers}. In greater detail, for \textit{OffensiveCommunity.net} and \textit{DarkWebMafias.net}, the first post of a thread has some differences in the HTML structure compared to other posts. In particular, the field of the post author is wrapped around some extra nodes that provide a special style to it. With the adoption of the parent XPath (\textbf{technique \#2}), it has been possible to infer a rule that works for every post's author. 
For \textit{Garage4Hackers.com} the XPath regex was not a sufficient approach to find all the post wrappers in a page. This is due to the fact that, when multiple resources are meant to be identified through an XPath regex, the XPath \texttt{//*[starts-with(@id, 'something')]} selector is used, which returns an array of nodes. Specifically, we were interested in nodes with id \texttt{post\_XXXXX}, but on the same page were also present nodes with id \texttt{post\_message\_XXXXX}, which caused the resolution of both content types. To overcome to that, identifying the required resources through a class was sufficient to solve the problem (\textbf{technique \#3}).
For all the forums, identifying the regex for the post wrapper is a special process that uses a \textbf{variant of technique \#2}, where the container is identified by incremental steps. Starting from  snippets of text from different posts of the page, a first XPath is calculated. Subsequently, with the user interaction, it is refined removing the unnecessary child until the whole post is correctly classified with the XPath calculated.
Moreover, for all the forums, inferring a stable XPath identifier for the next page buttons is not possible. This depends from the number of buttons inside of the navigation wrapper, which changes depending on the number of available pages or even when moving to the second page. To circumvent this problem, again a class comes in help \textbf{(technique \#3)}.
The double class (technique \#4) has never been used with the 4 forums analyzed. Nonetheless, it was proven to be necessary for another forum, a XenForo board, which was used benchmark for some first experiments.

\subsection{Network patterns and behaviour}

In order to evaluate how network traffic generated by \CARONTE\ compares w.r.t. network traffic generated by humans (i.e. legitimate users) and state-of-the-art crawlers, we performed an experiment employing the Amazon Mechanical Turk platform. This enables us to compare \CARONTE\ against both `undesirable' and `desirable' traffic from the perspective of the criminal forum administrator.

\subsubsection{Experiment methodology}

\label{sec:expmethod}

\paragraph{Human navigation experiment}

To generate human traffic towards our forums, we rely on Amazon Mechanical Turk (MTurk). From the literature review, we identify three main experimental variables characterizing the habits of a regular user on the Internet:

\begin{itemize}
    \item \textbf{Var1}: The interest raised in the reader by the content may lead him to read carefully all the content of a certain thread or not, resulting in skimming and moving quickly to a next resource~\cite{dupret2010model,guo2012beyond};
    \item \textbf{Var2}: The desire of privacy of the user, which may be high or low, resulting in the adoption of solutions that prevent JavaScript to be executed or not to avoid fingerprinting techniques~\cite{pastrana2018crimebb,allodi2017economic,torjs_thehackernews_2013};
    \item \textbf{Var3}: The propensity of an user to open several resources in parallel before actually browsing them or instead opening them one per time, reading their content first before moving to the next resource~\cite{dupret2008user}.
\end{itemize}    

To control for possible interdependencies between these dimensions, we create a $2^{3-1}$ \textit{fractional factorial experimental design}, that allows us to reduce the number of experimental conditions from eight to four~\cite{box19612}. The experimental treatments and design are reported in Table~\ref{tab:treatments}, and~\ref{tab:exps} respectively.

\begin{table}[t]
\small
\centering
\caption{Experimental features and treatments}
\label{tab:treatments}
\begin{tabular}{p{0.05\columnwidth} p{0.35\columnwidth} p{0.2\columnwidth} p{0.2\columnwidth}}
\toprule
\# & Exp. variable                                                                                                & Treatment A          & Treatment B             \\\midrule 
\textbf{Var1}          & The reader is interested in the content or skims a few posts&  Read all the content inside of the thread & Skim thread or read first post \\ 
\textbf{Var2}          & The user enables or disables JavaScript on Tor Browser       & Enabled              & Disabled                \\ 
\textbf{Var3}          & Opening resources in parallel or sequentially                & Sequential           & Parallel                \\ 
\bottomrule
\end{tabular}
\end{table}

\begin{table}[t]
\small
\centering
\caption{Treatment combination and experiments.}
\label{tab:exps}
\begin{tabular}{lllllllll}
\toprule
 & \multicolumn{2}{c}{Exp1} & \multicolumn{2}{c}{Exp2} & \multicolumn{2}{c}{Exp3} & \multicolumn{2}{c}{Exp4} \\ 
\cmidrule(lr){2-3}\cmidrule(lr){4-5}\cmidrule(lr){6-7}\cmidrule(lr){8-9}
   & A           & B           & A           & B           & A           & B           & A           & B           \\ \midrule
\textbf{Var1}           & -           & +           & -           & +           & +           & -           & +           & -           \\ 
\textbf{Var2}           & +           & -           & -           & +           & +           & -           & -           & +           \\ 
\textbf{Var3}           & +           & -           & -           & +           & -           & +           & +           & -           \\ 
\bottomrule
\end{tabular}
\end{table}

\subsubsection{Experimental design and setup}

\begin{figure}[t]
    \centering
    \includegraphics[width=0.6\columnwidth]{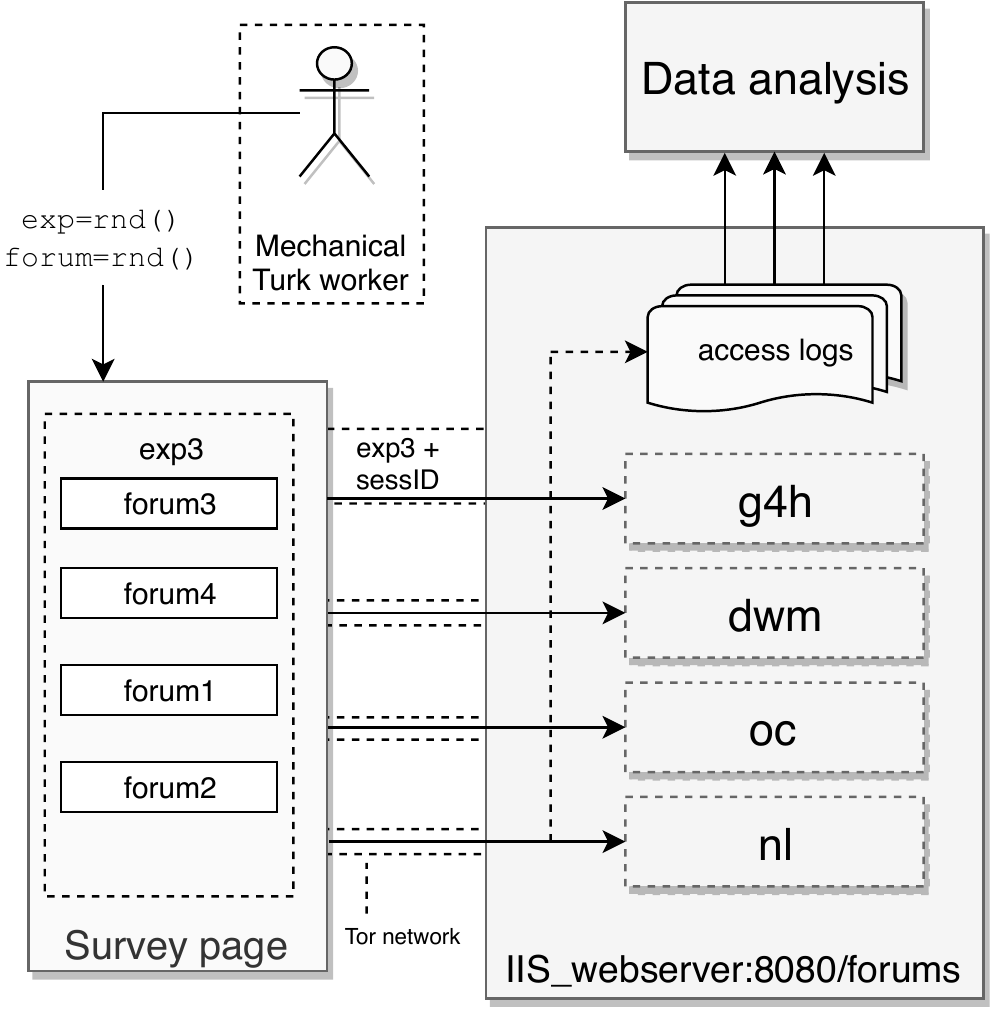}
    \begin{minipage}{0.9\columnwidth}
    \footnotesize
    The forums are deployed on an internal system at the University. Resources are accessed by industry standard automated tools (scrapers), \CARONTE, and MTurks. All tools access the local resources through the TOR network. Each MTurk is randomly assigned to an experiment setup with different conditions (see Table~\ref{tab:exps}). Internal network logs allow us to backtrack user requests to specific experimental setups.
    \end{minipage}
    \caption{Experimental setup}
    \label{fig:setup}
\end{figure}

An overview of the experimental setup is shown in Figure \ref{fig:setup}.
The setup implementation  has been carried out in three stages: the selected web forums (ref. Table~\ref{tab:forums}) are hosted on an IIS web server (vers. 10) where access logging is enabled. We prepare an Amazon Mechanical Turk task reflecting the experimental design (ref. Table~\ref{tab:exps}). The task includes eight questions based on the content of the forum webpages (two multiple-choice questions per forum).
The task included detailed step-by-step instructions that respondents had to follow. Such instructions serve the purpose of enforcing the treatment in the experiment; for example, \emph{Exp3} requires users to read all content of a thread (\textbf{Var1, A}), have JavaScript enabled (\textbf{Var2, A}), and open forum tabs in parallel (\textbf{Var3, B}): 
\begin{quote}
\emph{[...] open in separate tabs all threads you think are relevant to those two topics (\textbf{Var3, B}).\\
While reading the forum threads, please also skim through to at least the second thread page (\textbf{Var1, A}), if present, and even if you already found the answer.}
\end{quote}
Notice that, as JavaScript is enabled by default in TOR browser, there is no instruction for \textbf{Var2, A}.
 When a respondent accesses the task on AMT, he or she is assigned randomly to an experimental condition. Further, each instance of the experiment randomizes the forum order to minimize cross-over effects. The last step consists of enabling us to collect the generated network requests. To avoid limitations imposed by the TOR circuit refresh mechanism\footnote{\texttt{MaxCircuitDirtiness} - https://www.torproject.org/docs/tor-manual-dev.html.en} that may change the IP address of users every 10 minutes, we set a cookie on the user's browser with a unique session ID. 
We use the same strategy to track the experimental condition to which the user has been randomly assigned to at access time.

\subsubsection{Results}

\begin{figure*}[t]
\centering
\includegraphics[width=0.31\textwidth]{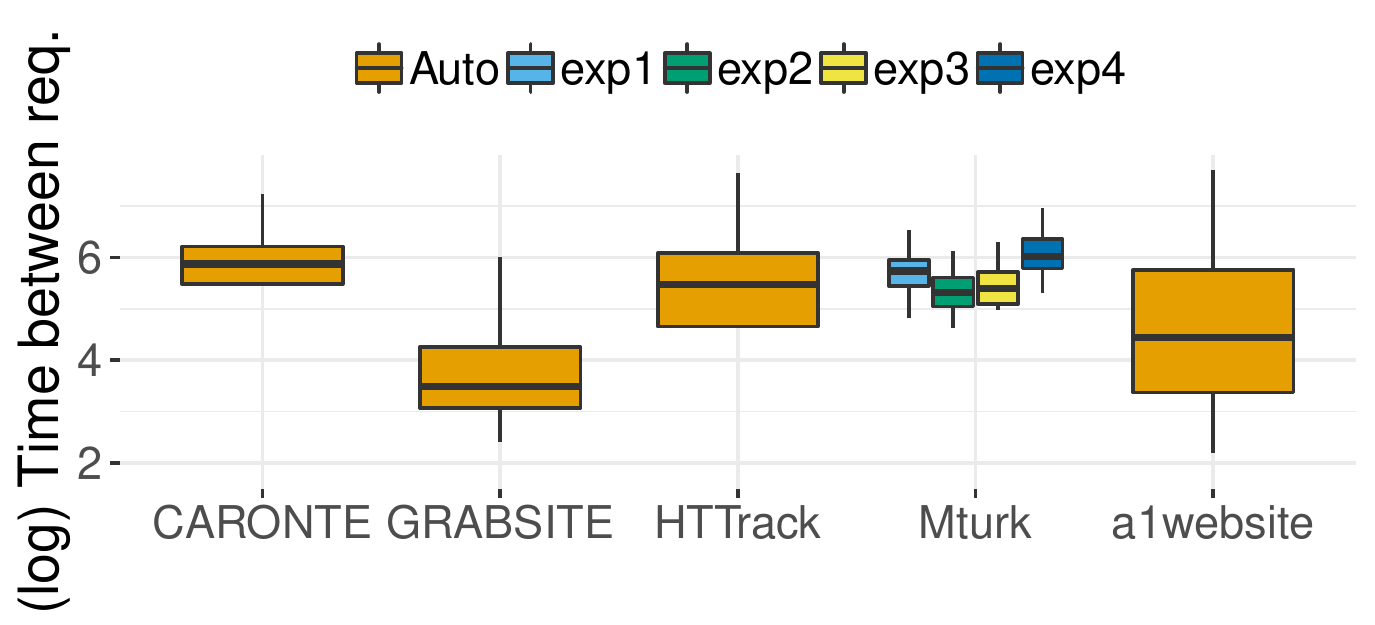}
\includegraphics[width=0.31\textwidth]{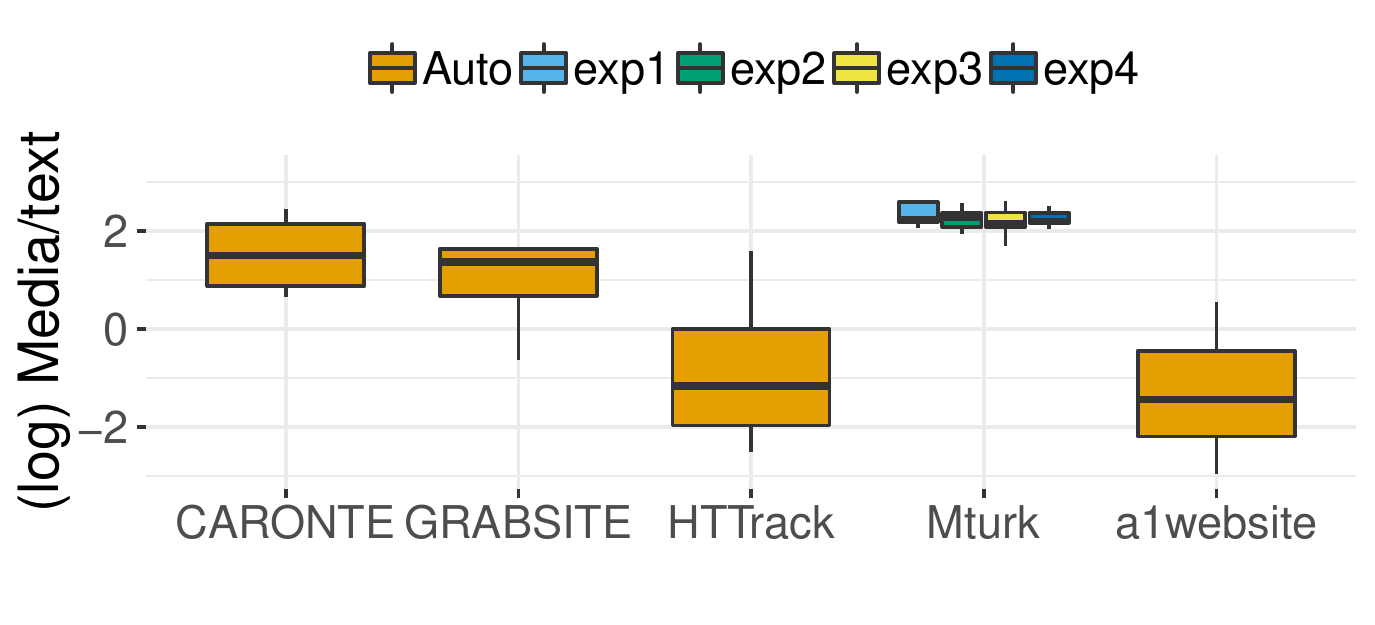}
\includegraphics[width=0.31\textwidth]{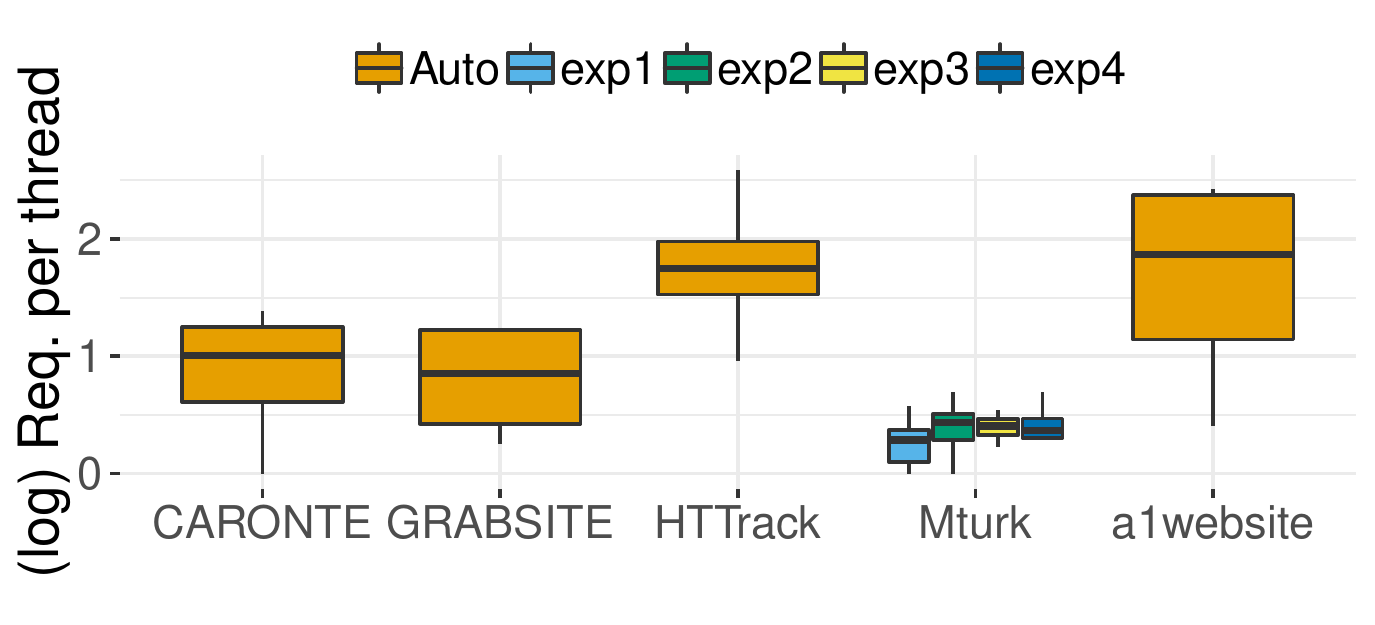}
\caption{Evaluation of \CARONTE\ against state-of-art tools and MTurks}
\label{fig:eval}
\end{figure*}

Figure~\ref{fig:eval} 
reports the network analysis for \CARONTE\ compared to the state of the art tools and the MTurks. As emerged from the requirements, \CARONTE{} should not exhibit greed in resource fetching, but should access them with lower frequency, accordingly to the resource content. On one hand, if is true that avoiding request bursts is probably more than enough against an automated monitoring tool, on the other hand it guarantees an extra layer of stealthiness in case of human verification and could fool ML-based robot-detection systems. Therefore, we compare this traffic throttling model with humans and other tools by monitoring the amount of requests per thread and the time between them. From the comparisons, emerges that the time elapsed between two different requests\footnote{\label{requests}A \textbf{request} refers to all the calls to a page of a thread, without considering all the linked content downloaded.} produced by humans is comparable to \CARONTE{}'s and HTTrack, while the others perform more aggressively. 
For what concerns the media/text ratio of the sessions, \CARONTE\, together with \grabsite, perform quite close to humans. 
Finally, we've compared the number of requests issued per thread\footnote{\textbf{Requests per thread} refer to the set of all the \textit{requests} that have been fired by an actor inside of each and every thread.}: \CARONTE\ and \grabsite\ perform again better when compared to humans than the other two tools, but their behavior slightly differs from MTurks. Overall, we observe that \CARONTE\ network trace is consistently very similar to human-generated network traffic, whereas other tools are clearly different over one or more dimensions.

\section{Discussion}
\label{sec:discussion}

\CARONTE's training module proved effective in flexibly learning diverse forum structures. Differently from ML-based systems, the adopted semi-automated procedure allows the tool to reliably identify relevant structures in the DOM of a page, while avoiding entirely the need to collect massive amounts of pre-existent data (for the training and validation) that might jeopardize the researcher activity. Whereas this does come at the price of additional human-sourced work w.r.t. fully-automated procedures, \CARONTE\ is meant to be employed over (the few) highly-prominent underground communities where the threat model \CARONTE\ addresses is realistic. The presented proof-of-concept has been tested over four diverse forum structures, and can be expanded in future work beyond the `forum' domain (e.g. e-commerce criminal websites).

From the network analysis it emerges that \CARONTE\ reproduces coherently the three investigated features when compared to humans and performs better, on average, than the other tools. Our tool produces the multimedia traffic of a regular human actor, together with \grabsite{}, while the other two tools diverge from this behavior; we suspect that this is to be traced back to some optimization mechanisms which avoid to reissue requests for the same resource, without even issuing an HEAD HTTP request. A regular browser instead will always reissue the request while loading another page, if not explicitly instructed by a server-side caching policy. Nonetheless, we have found no confirmation in the documentation of these tools. 
With regards to the number of requests generated per thread, there's a noticeable difference when compared to humans. This is probably caused by MTurks skipping some pages in the threads. In fact, in multiple cases, the downloaded forums have plenty of `useless' replies to threads, which may result in a decreased interest from the reader, possibly leading to skipping the following pages. The observed difference in generated requests per thread between \CARONTE\ and \grabsite\ and the other two tools is caused by the fact that they follow also non relevant links, such as content re-displacement in the page. In particular, this last behavior represent a well-known traffic feature of a crawler. To improve this, it could be possible to instruct our tool to ignore threads where content is redundant and extremely short.

As mentioned in the \textit{network patterns and behavior} subsection, we have monitored some extra features that may represent a red-flag in crawler detection. Nonetheless, they're not part of the experiment since are enforced conditions (like the filling of the referral field) by design of our tool. These are shown, for reference, in Table~\ref{tab:flags}: \CARONTE\ explores threads one at a time, sequentially, while other crawlers tend to open multiple resources in parallel. \texttt{A1 Website Download} has never filled the referrer URL in the HTTP requests, highlighting the fact that this request has not been sent from a legitimate browser. In the last analysis, for these monitored aspects, we can say that \texttt{HTTrack} performs better than the others in terms of browser features exhibited. 

Currently, \CARONTE{} is limited to the circumvention of passive traffic monitoring and intrusion detection tools; a more sophisticated or active adversary could introduce new techniques to recognize if the user connected is a human (live chat, custom made CAPTCHAs, ...) or could adopt refined anti-crawling mechanisms that leverage on ML.

\begin{table}[t]
\small
\centering
\caption{Extra features monitored.}
\label{tab:flags}
\begin{tabular}{p{0.2\columnwidth}p{0.05\columnwidth}p{0.06\columnwidth}p{0.08\columnwidth}p{0.13\columnwidth}p{0.14\columnwidth}}
\toprule
Tool                         & JS & Styles & Cache & Seq./Par. & Referrals \\ \midrule
\CARONTE\                    & \ding{55}             & \ding{51}         & \ding{51} & Seq.          & \ding{51}            \\ 
\grabsite                    & \ding{51}             & \ding{51}         & \ding{55} & Par.            & \ding{51}            \\ 
\texttt{HTTrack}             & \ding{51}             & \ding{51}         & \ding{51} & Par.            & \ding{51}            \\ 
\texttt{A1Website} & \ding{51}             & \ding{51}         & \ding{55} & Par.            & \ding{55}            \\ 
\bottomrule
\end{tabular}
\end{table}

\section{Conclusions, ethical aspects and future work}
\label{sec:conclusions}
Automated tools that gather data in a stealthy way from high-profile forums are a growing need in our society, due to the increase of cyberattacks and the amount of data in these platforms. \CARONTE{} has proven to be a flexible and suitable solution for stealthy information gathering from cybercriminal forums, but it can be adopted disregard the content of the platform explored. This could allow collecting data from virtually any forum for several purposes, such as bullying monitoring, suicidal prevention and terrorism monitoring, with the aid of NLP modules. Nonetheless, it may become an instrument for investigations driven by nations that apply strong censorship and violent oppression, jeopardizing the freedom of whistle-blower reporters. 

As future work, we plan to introduce support for solving CAPTCHAs, which represent a barrier for \CARONTE{} as well as considering website fingerprinting to infer browser-based monitoring capabilities. 

\CARONTE{} will be made available to interested researchers.
\bibliographystyle{acm}
\bibliography{biblio,bib,short-names,security-common}

\end{document}